\documentclass[final]{ustcstep}

\usepackage[dvips]{graphicx}
\usepackage[square]{natbib}

\def\gsim{\;\lower4pt\hbox{${\buildrel\displaystyle >\over\sim}$}\;}
\def\lsim{\;\lower4pt\hbox{${\buildrel\displaystyle <\over\sim}$}\;}
\def\grls{\;\lower4pt\hbox{${\buildrel\displaystyle >\over <}$}\;}

\newcommand\addr[2]{{\footnotesize \it $^{#1}$#2}\\}

\begin{document}

\title{Deflected propagation of a coronal mass ejection from the corona to interplanetary space}

\author{Yuming Wang,$^{1,2,*}$ Boyi Wang,$^1$ Chenglong Shen,$^{1,2}$ Fang Shen$^3$ and No\'e Lugaz$^4$\\[1pt]
\addr{1}{CAS Key Laboratory of Geospace Environment, Department
of Geophysics and Planetary Sciences, University of Science and}
\addr{ }{ Technology of China, Hefei, Anhui 230026, China}
\addr{2}{Synergetic Innovation Center of Quantum Information and
Quantum Physics, University of Science and Technology of China,}
\addr{ }{ Hefei, Anhui 230026, China}
\addr{3}{SIGMA Weather Group, State Key Laboratory of Space
Weather, Center for Space Science and Applied Research, Chinese}
\addr{ }{ Academy of Sciences, Beijing 100190, China}
\addr{4}{Space Science Center and Department of Physics,
University of New Hampshire, Durham, NH, USA}
\addr{*}{Corresponding Author, Contact: ymwang@ustc.edu.cn}}

\maketitle
\tableofcontents

\begin{abstract}
Among various factors affecting the space weather effects of a
coronal mass ejection (CME), its propagation trajectory in the
interplanetary space is an important one determining whether and
when the CME will hit the Earth. Many direct observations have
revealed that a CME may not propagate along a straight trajectory in
the corona, but whether or not a CME also experiences a deflected
propagation in the interplanetary space is a question, which has
never been fully answered. Here by investigating the propagation
process of an isolated CME from the corona to interplanetary space
during 2008 September 12 -- 19, we present solid evidence that
the CME was deflected not only in the corona but also in the
interplanetary space. The deflection angle in the interplanetary
space is more than $20^\circ$ toward the west, resulting a
significant change in the probability the CME encounters the Earth. A
further modeling and simulation-based analysis suggest that the
cause of the deflection in the interplanetary space is the
interaction between the CME and the solar wind, which is different
from that happening in the corona.
\end{abstract}


\section{Introduction}
Coronal mass ejections (CMEs) that originate from solar source
regions facing Earth are thought to be one of the main drivers of
hazardous space weather. Such CMEs usually appear like a halo in a
coronagraph. However, not all of such halo CMEs hit the Earth. Only
about 60\%--70\% of front-side halo CMEs are found to be associated
with an ejecta near the Earth, and the fraction is even smaller,
$\sim50\%$, for geoeffective front-side CMEs
\citep[e.g.,][]{Webb_etal_1996, Webb_etal_2001, Cane_etal_1998a,
Plunkett_etal_2001, Wang_etal_2002a, Berdichevsky_etal_2002,
Yermolaev_etal_2005}. On the other hand, CMEs originating from solar
limb are possible to hit the Earth \citep[e.g.,][]{Webb_etal_2000,
Zhang_etal_2003, Cid_etal_2012}. Such events might cause so called
`problem storms', which cannot be found any associated on-disk CMEs
\citep{Webb_etal_2000, Schwenn_etal_2005, Zhang_etal_2007}.
Statistical studies suggested that the association of ejecta to CMEs
is about 60\% \citep[e.g.,][]{Lindsay_etal_1999, Cane_etal_2000,
Cane_Richardson_2003}.

For problem storms, there are several possible explanations. One of
them is that such storms are caused by CMEs with a large
longitudinal extension, which could sweep through the Earth even if
originating far from the disk center \citep{Webb_etal_2000,
Zhang_etal_2003}. The existence of stealth CMEs, which have been
recently observed by the Solar-Terrestrial Relationship
Observatories (STEREO, \citealt{Kaiser_etal_2008}), is another
hypothesis. Such CMEs do not leave any footprints behind them in EUV
observations \citep{Robbrecht_etal_2009} though they may face to the
observer. Statistical studies suggested that stealth CMEs are not a
rare phenomenon, but may correspond to one third of all front-side
CMEs \citep{Ma_etal_2010, Wang_etal_2011}. Both of the above
explanations could explain the problem storms but cannot explain why
some CMEs originating from the solar disk center miss the Earth.

A promising explanation is that CMEs may be deflected during their
propagation in the corona and interplanetary space. CME-CME
interaction is a cause of CME deflection \citep{Wang_etal_2011,
Shen_etal_2012, Lugaz_etal_2012}. The deflected propagation angle
could be ten degrees or even larger. More interestingly,
observations imply that, even if there was only one CME, it could be
deflected by the background solar wind and magnetic field. The
deflection of isolated CMEs in the plane-of-sky in corona was
reported since 1986 \citep{MacQueen_etal_1986}, and has been studied
by many researchers \citep[e.g.,][]{Gopalswamy_etal_2003,
Gopalswamy_etal_2004, Gopalswamy_etal_2009, Cremades_Bothmer_2004,
Cremades_etal_2006, Wang_etal_2011, Lugaz_etal_2011, Kahler_etal_2012, Zuccarello_etal_2012, 
Yang_etal_2012, DeForest_etal_2013, Zhou_Feng_2013}. By using
STEREO data, it is found that the CME deflection in corona could be
more than $20^\circ$, and appear to be controlled by the gradient of
the corona magnetic energy density \citep{Shen_etal_2011,
Gui_etal_2011}.

Whether or not an isolated CME could be also deflected in the
interplanetary space is an open question, because the state of
interplanetary space in some sense is much different from that of
the corona. In the corona, the magnetic field is dominant, and the
solar wind has not been well developed; whereas in the
interplanetary space, the solar wind becomes dominant, as magnetic
fields decrease with distance and the solar wind has fully
accelerated.

The idea of CME deflection in interplanetary space was first
proposed by \citet{Wang_etal_2002a} through a statistical study, and
then developed in their follow-up works \citep{Wang_etal_2004b,
Wang_etal_2006a}. For a CME faster than the ambient solar wind, the
interplanetary magnetic field will be piled up ahead of the CME and
cause an eastward deflection; while for a CME slower than the solar
wind, the magnetic field will accumulate behind the CME and cause a
westward deflection \citep[see Fig.4 of][]{Wang_etal_2004b}. The
deflection angle could reach tens degrees. But the direct evidence
of such a deflection has rarely been reported. Some recent case
studies linking remote-sensing and in-situ data indirectly suggested
that CMEs are possible to be deflected in interplanetary space
\citep[e.g.][]{Kilpua_etal_2009, Rodriguez_etal_2011,
Isavnin_etal_2013}. With the aid of triangulation method, the
propagation direction of CMEs in the heliosphere was investigated by
\citet{Lugaz_2010} based on STEREO observations. He found that 6 out
of 13 CMEs perhaps experienced a deflected propagation with the
deflection angle larger than $20^\circ$, and both eastward and
westward deflections exist. That previous work was focused on
the development of new analysis techniques, therefore, the precise
deflection process and its possible cause was not analyzed.

In this study, we present the first detailed analysis of the
deflection of an isolated CME during its heliospheric propagation.
We focus on a CME which occurred on 2008 September 12 and study its
trajectory as well as the physical causes and mechanisms for the
observed deflection. The observations of this event will be
presented in the next section. In Sec.~\ref{sec_pro}, by applying a
variety of models, we will study the propagation of the CME,
including the evolution of its velocity and direction. We draw
conclusions in section 4 and discuss our results in term of physical
causes of CME deflection in section 5.

\section{Observations}\label{sec_obs}

\subsection{Instruments and data}
\begin{figure}[tb]
  \centering
  \includegraphics[width=1.\hsize]{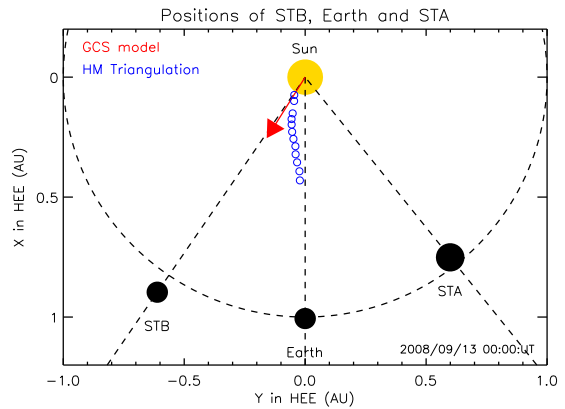}
  \caption{Positions of STA and STB relative to the Earth near which SOHO locates at the beginning of
September 13. The red arrow denotes the initial propagation
direction of the CME derived by GCS model. The blue circles indicate
the propagation direction of the CME from out corona to
interplanetary space, which are inferred from STEREO imaging data by
HM triangulation method. The curved path formed by the blue circles
suggests that the CME experienced a deflection
process.}\label{fg_pos}
\end{figure}

The imaging data used in the following analysis are from the
Large Angle and Spectrometric Coronagraph (LASCO,
\citealt{Brueckner_etal_1995}) onboard SOHO spacecraft and the Sun
Earth Connection Coronal and Heliospheric Investigation (SECCHI)
suites \citep{Howard_etal_2008} onboard both STEREO-A (STA) and
STEREO-B (STB) spacecraft. The in-situ data of interplanetary
magnetic field and solar wind plasma at 1 AU are from
IMPACT~\citep{Acuna_etal_2008} and PLASTIC~\citep{Galvin_etal_2008}
instruments onboard STA and STB spacecraft and
MFI~\citep{Lepping_etal_1995}, SWE~\citep{Ogilvie_etal_1995} and
3DP~\citep{Lin_etal_1995} instruments onboard Wind Spacecraft. SOHO
and Wind is located at the first Lagrange point of the Sun-Earth
system, and the STEREO twin spacecraft fly in Earth's orbit with an
increasing separation to the Earth. The positions of STA and STB in
HEE coordinates at the beginning of 2008 September 13 are plotted in
Figure~\ref{fg_pos}. At that time, STA is separated away from the
Earth by about $39^\circ$, and STB by about $34^\circ$. The LASCO
instrument carries two working cameras, C2 and C3, that covering the
corona from 2.0 -- 30 $R_S$. In SECCHI suites, there are cameras,
COR1, COR2, HI1 and HI2, monitoring the corona and interplanetary
space from 1.4 $R_S$ to beyond 1 AU. These imagers provide seamless
observations of the kinematic evolution of a CME from multiple
angles of views.

\subsection{Imaging observations}\label{sec_imaging}
\begin{figure*}[!tbh]
  \centering
  \includegraphics[width=1.\hsize]{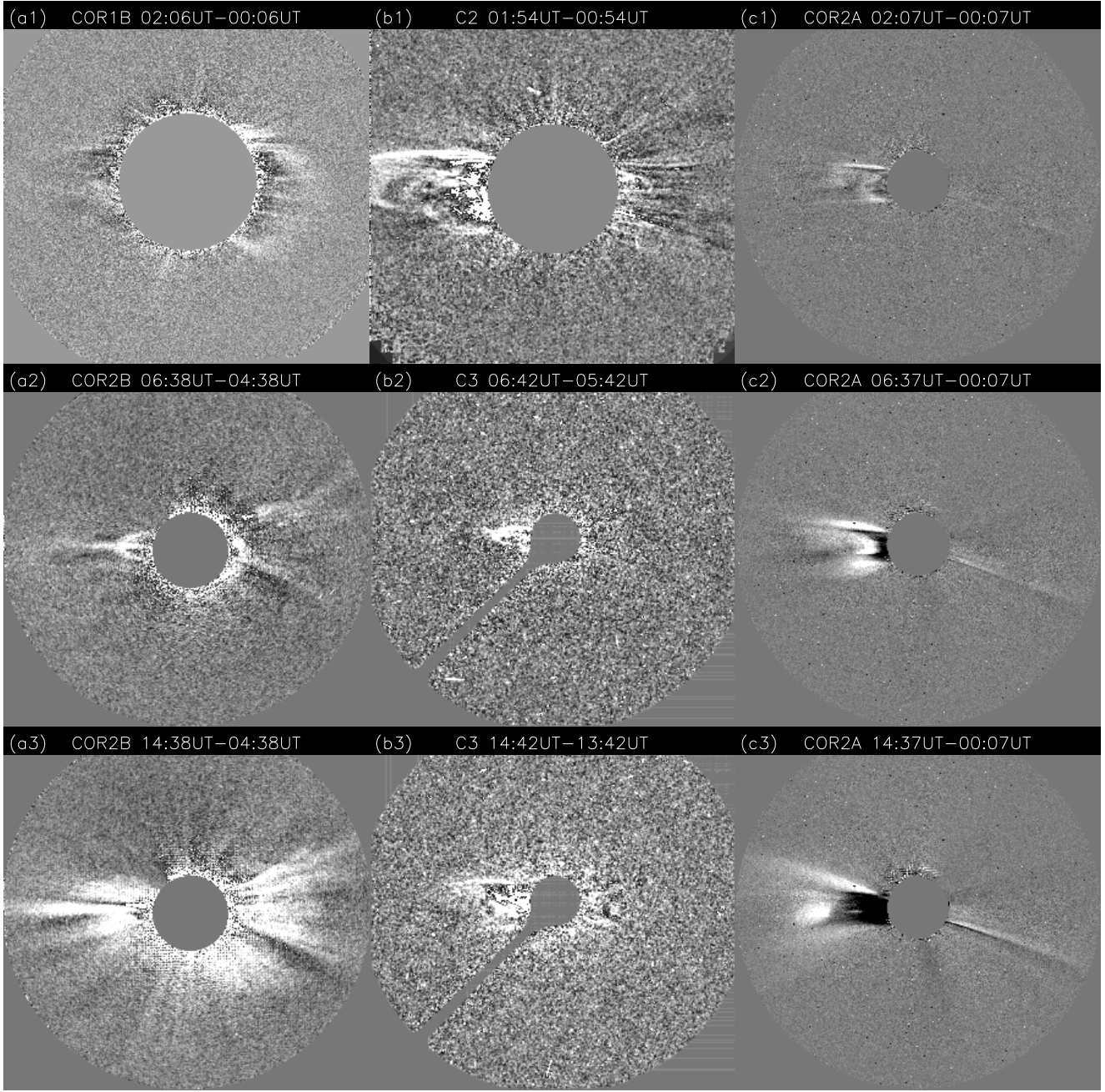}
  \caption{Snapshots of the CME taken by STB/COR (panel a), SOHO/LASCO (panel
b) and STA/COR (panel c) on September 13. In both views of SOHO and
STA, the CME looks like a east-limb event. But in the view of STB,
it is a partial halo CME. Three movies generated from difference images
have been attached in the online material to show the propagation of the CME 
viewed by SOHO, STA and STB, respectively.}\label{fg_cme}
\end{figure*}

The CME is a slow one as observed by STEREO and SOHO spacecraft (see
Fig.~\ref{fg_cme}, and associated movies). It roughly traveled
in the ecliptic plane. In both the views of STA and SOHO, the CME
looks like an east-limb event. Due to data gap, its first appearance
in the field of view (FOV) of STA/COR1 is not clear, but must be
before 15:00 UT on September 12, when the CME was already in the FOV
of COR2. SOHO/LASCO C2 camera also captured an eastward CME starting
at 15:30 UT, and 8 hours later, the CME appeared in the FOV of LASCO
C3. Differently, in the view of STB, the CME presented a halo shape
with main part expanding toward the west. It appeared in the FOV of
STB/COR1 at about 19:38 UT on September 12, and emerged into the FOV
of STB/COR2 about 9 hours later. Since it is a very slow CME, the
CME fully developed into the FOVs of all the three coronagraphs
around the beginning of September 13. Thus we just focus on the
dynamic evolution of the CME since that time.


\begin{figure*}[tbh]
  \centering
  \includegraphics[width=\hsize, angle=0]{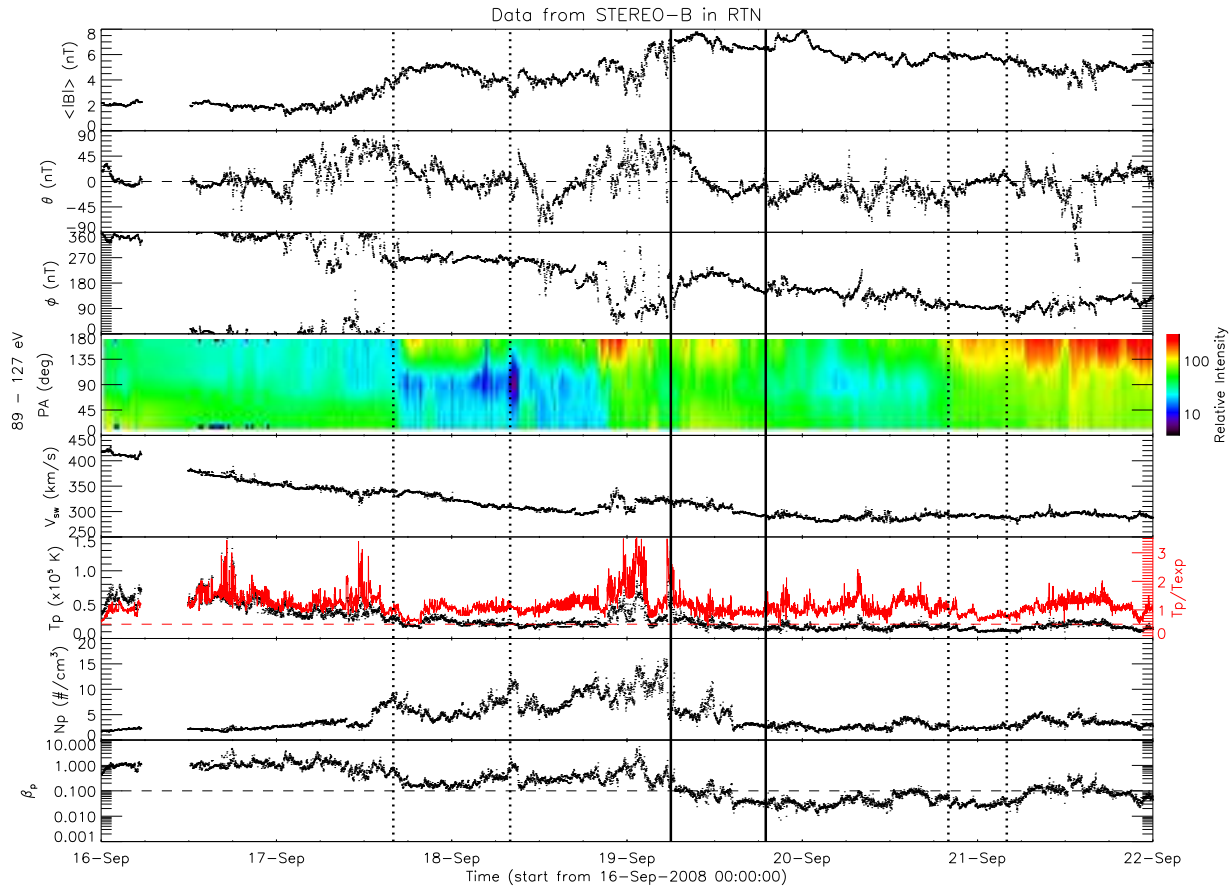}
  \caption{Interplanetary magnetic field and solar wind plasma data at 1 AU
from in-situ instruments on board STB. The panels from the top to
bottom are the magnetic field strength ($<|B|>$), elevation
($\theta$) and azimuthal ($\phi$) angles of the magnetic field
direction, the electron pitch-angle (PA), the solar wind bulk speed
($v_{sw}$), the proton temperature ($T_p$), number density ($N_p$)
and beta ($\beta$). The red curve in the 7th panel is the ratio of
proton temperature to the expected proton temperature ($T_{exp}$),
which is calculated based on the empirical formula by
\citet{Lopez_Freeman_1986}.}\label{fg_stb}
\end{figure*}

The CME appears as an east-limb event for both STA and SOHO.
Considering the positions of STA and SOHO (Fig.\ref{fg_pos}), it
suggests that the initial CME direction of propagation, i.e., within
15 $R_S$, is on the east-side of the Sun-Earth line. For STB, the
CME obviously inclined to the west, suggesting that the CME initial
direction must be on the west-side of the Sun-STB line.
Furthermore, considering that the CME looks more likely halo in the
view of STB than in the view of SOHO, we may conclude that the
initial propagation direction of the CME is located between the
Sun-Earth line and the Sun-STB line and much closer to the later.
So far, we have the
first impression that the CME will encounter STB with its main body
and sweep the Earth with its flank. To verify this, we check
in-situ measurements from STA, STB and Wind.

\subsection{In-situ observations}\label{sec_insitu}
\begin{figure*}[t]
  \centering
  \includegraphics[width=\hsize, angle=0]{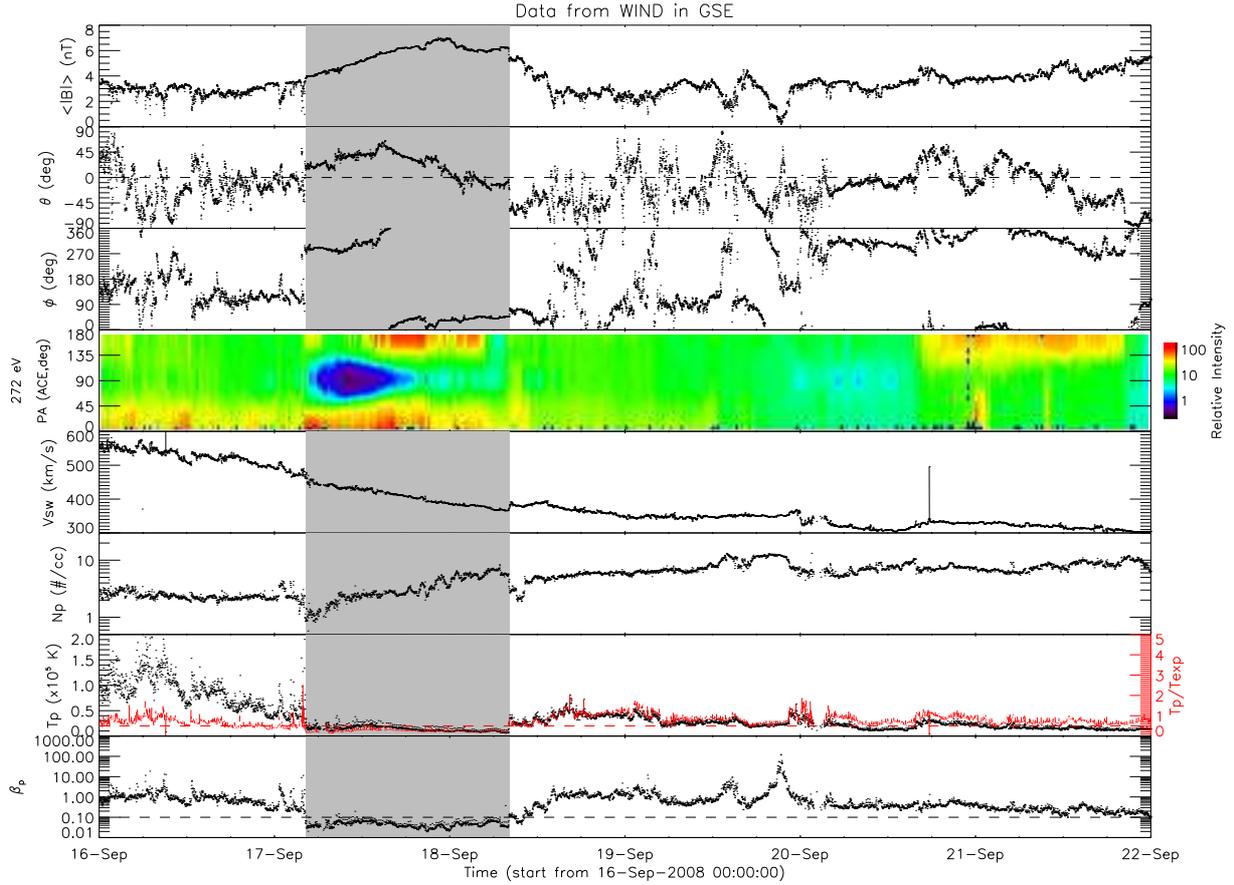}
  \caption{Same as Fig.~\ref{fg_stb} but from in-situ instruments on board Wind. The
shadow region indicates an MC.}\label{fg_wind}
\end{figure*}

\begin{figure*}[t]
  \centering
  \includegraphics[width=\hsize, angle=0]{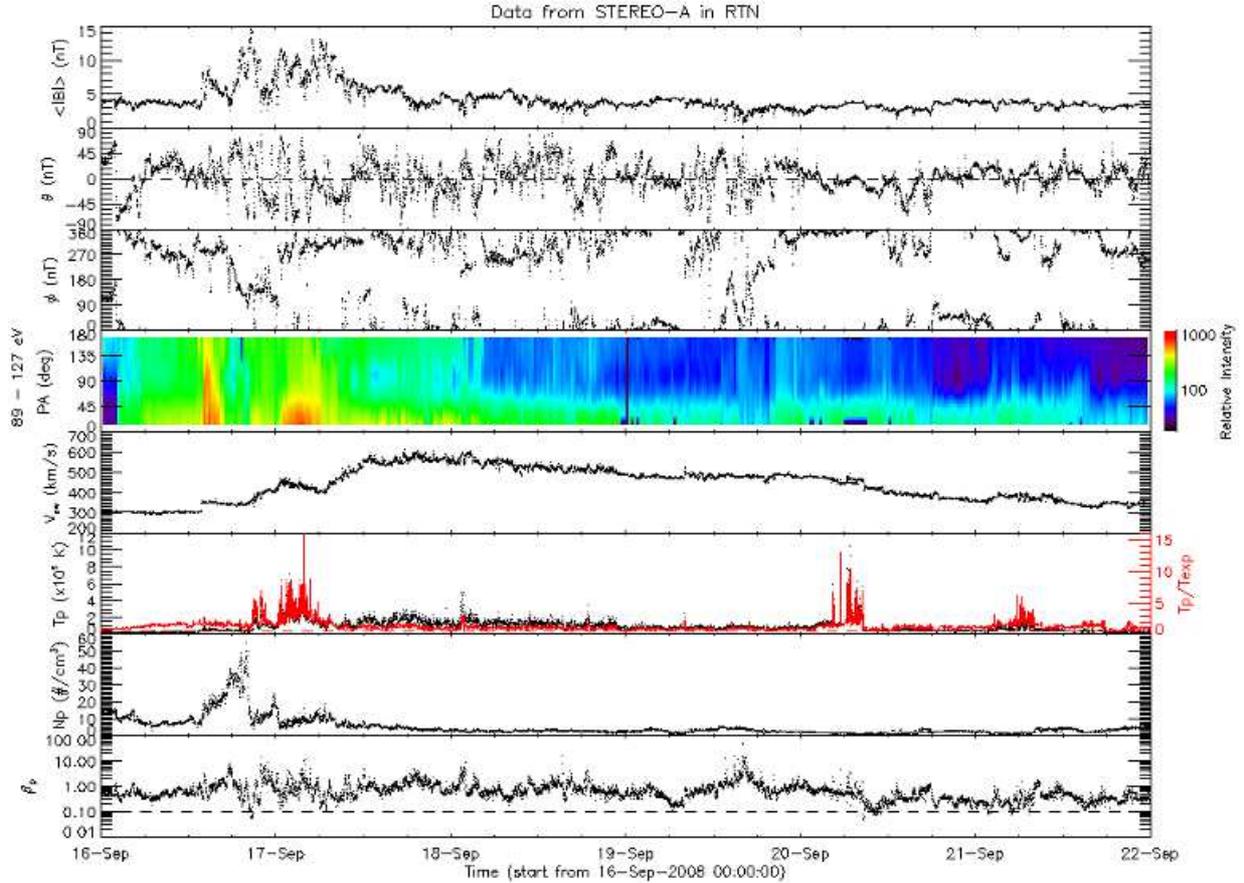}
  \caption{Same as Fig.~\ref{fg_sta} but from in-situ instruments on board STA.}\label{fg_sta}
\end{figure*}

According to the coronagraph observations, we find that the initial
speed of the CME is slower than 300 km s$^{-1}$. Even if the
acceleration by the solar wind is taken into account, its average
transit speed will not be larger than 500 km s$^{-1}$, which is
higher than the typical value for the background solar wind speed.
Thus it is expected that the interplanetary counterpart of the CME
should be observed at 1 AU at least three days later. We examine the
in-situ data from STA, STB and Wind spacecraft for 6 days starting
from September 16. The parameters of interplanetary magnetic field
and solar wind plasma during this period at three observational
points are presented in Figure~\ref{fg_stb} through \ref{fg_sta}.

In Wind data, we can identify one and only one interplanetary CME during
September 17 -- 18 (see the shadow region in Fig.\ref{fg_wind}),
which is a magnetic cloud (MC) following, e.g., \citet{Burlaga_etal_1981}'s
definition. Its front boundary is at about 04:20 UT on September 17
and the rear boundary at about 08:00 UT on the next day. All the MC
signatures are very clear, which basically include (1) the enhanced
magnetic field strength, (2) large and smooth rotation of magnetic
field direction, (3) declining profile of solar wind velocity, (4)
bi-directional streaming of supra-thermal electrons, (5) low proton
temperature and (6) low proton $\beta$ ($<0.1$ generally)
\citep[e.g.,][]{Burlaga_etal_1981, Gosling_etal_1987,
Lepping_etal_1990, Farrugia_etal_1993a, Richardson_Cane_1995}. The
average speed of solar wind during the MC is about 415 km s$^{-1}$,
suggesting the corresponding CME lifting off from the Sun around the
beginning of September 13. Particularly, the STEREO imaging data
show that there are no other CMEs directing to the Earth within 2
days before and after the CME. Thus it is conclusive that the MC
observed by Wind is the interplanetary counterpart of the CME of
interest.


\begin{figure*}[!tbh]
  \centering
  \includegraphics[width=\hsize]{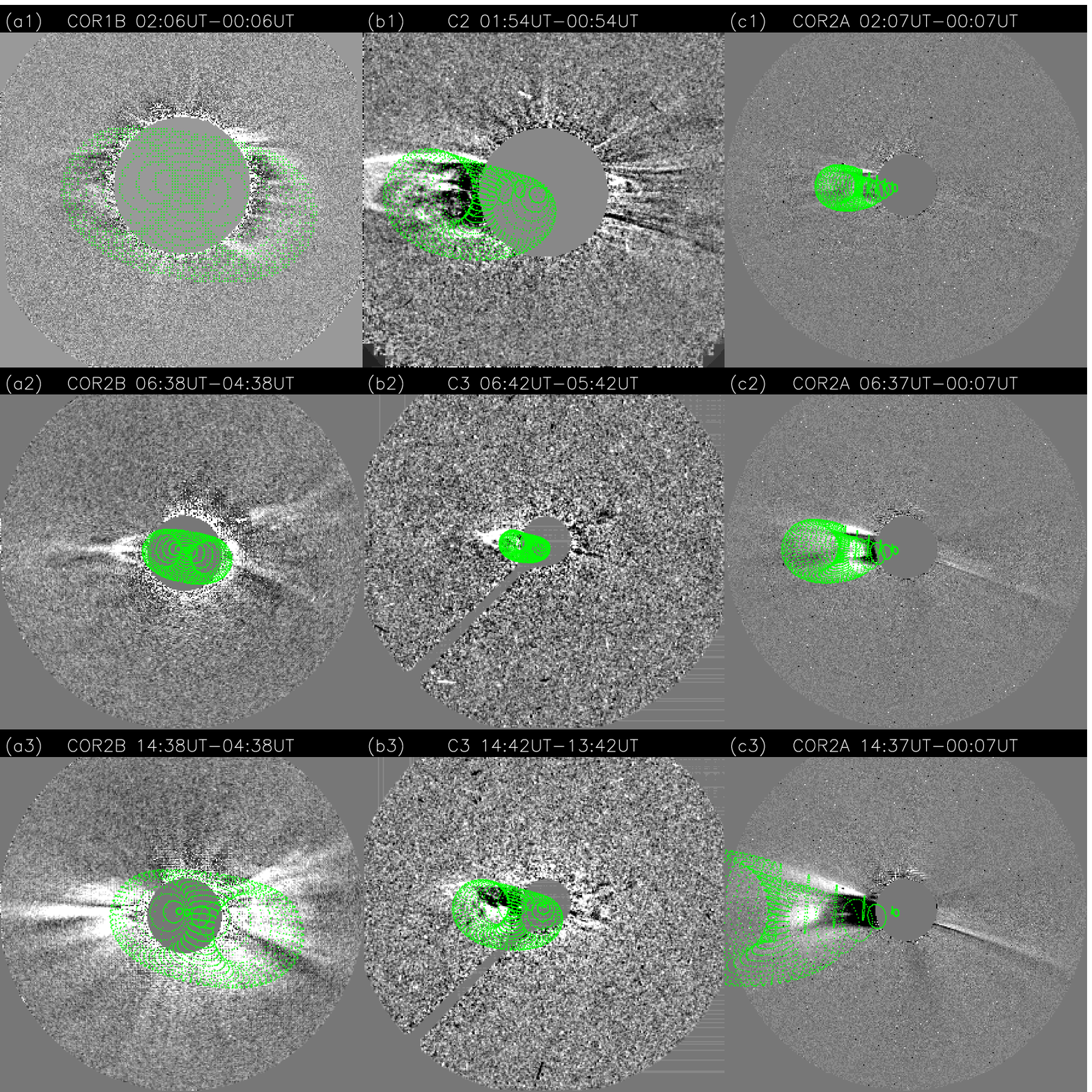}
  \caption{Coronagraph images showing the results of GCS model. The panels are the same as those
in Fig.~\ref{fg_cme} but with flux rope meshes superimposed.}\label{fg_gcs}
\end{figure*}

During the same period, STB did not capture any region
with clear signatures of an MC, but it is possible to identify
several MC-like or non-MC ejecta, e.g., the intervals between
September 17 16:00 UT and September 18 08:00 UT, between September
19 06:00 UT and 19:00 UT, and between September 20 20:00 UT and
September 21 04:00 UT (as indicated by vertical lines in
Fig.\ref{fg_stb}). None of them satisfies all the
previously-mentioned six signatures of a typical MC. In the the
first interval, there are only two signatures satisfied, i.e., the
magnetic field is stronger than that of ambient solar wind and solar
wind speed is declined. The second interval matches the most
signatures, including smooth rotation of magnetic field vector,
declining solar wind speed, bi-directional streaming of electrons
and low proton $\beta$, but does not have significantly enhanced
magnetic field and low proton temperature. The third interval also
satisfies only two signatures, which are smooth rotation of magnetic
field vector and low proton $\beta$. Considering that there was no
other CME roughly toward STB in the week starting from September 12,
we choose the ejecta in the second interval, which has the most
signatures of an MC, as the counterpart of the CME of interest. This
ejecta lost some signature of an MC is probably because the CME's
flank glanced over STB. The CME totally missed STA, as the STA data
shows typical solar wind at all times except during the period from
September 16 14:00 UT to September 17 12:00 UT, during which a
corotating interaction region (CIR) was formed between a high speed solar
wind stream and a low speed solar wind stream (Fig.\ref{fg_sta}).

One may notice that the CME's arrival at STB is about 2 days later
than that at Wind. Such a long delay could be attributed to the
curved front of the CME. A more detailed discussion on this issue
will be given in Sec.\ref{sec_con}. The analysis of imaging data
about the CME propagation in the corona has suggested that the CME's
main body should pass over STB and its flank glanced over the Earth,
but the in-situ data from multiple points at 1 AU reveal that the
fact is reversed, the CME's main body passed through the Earth and
its flank may have glanced over STB. This result suggests that the
CME be deflected during its journey from the corona to 1 AU.

\section{Propagation process}\label{sec_pro}
\subsection{In corona}\label{sec_corona}
CMEs are believed to have a flux rope
topology~\citep[e.g.,][]{Vourlidas_etal_2013}. Thus, the kinematic
process of the CME is studied by applying a forwarding modeling
with the aid of GCS model \citep{Thernisien_etal_2009,
Thernisien_2011}, which assumes that a CME has a flux-rope shape and
expands self-similarly. It uses six free parameters to shape
the flux rope, which are equivalent to height or heliocentric
distance of the leading edge, latitude and longitude of the
propagation direction, face-on and edge-on angular widths and tilt
angle of the main axis of the flux rope. We get these parameters of
a CME by fitting the GCS model to the observed outlines of a CME
viewed from all the angles of views of SOHO, STA and STB. This
model has been successfully applied to numerous CME events to study
the deflected propagation of CMEs by \citet{Gui_etal_2011}. One
may refer to the above references and therein for more details.

\begin{figure}[tbh]
  \centering
  \includegraphics[width=\hsize]{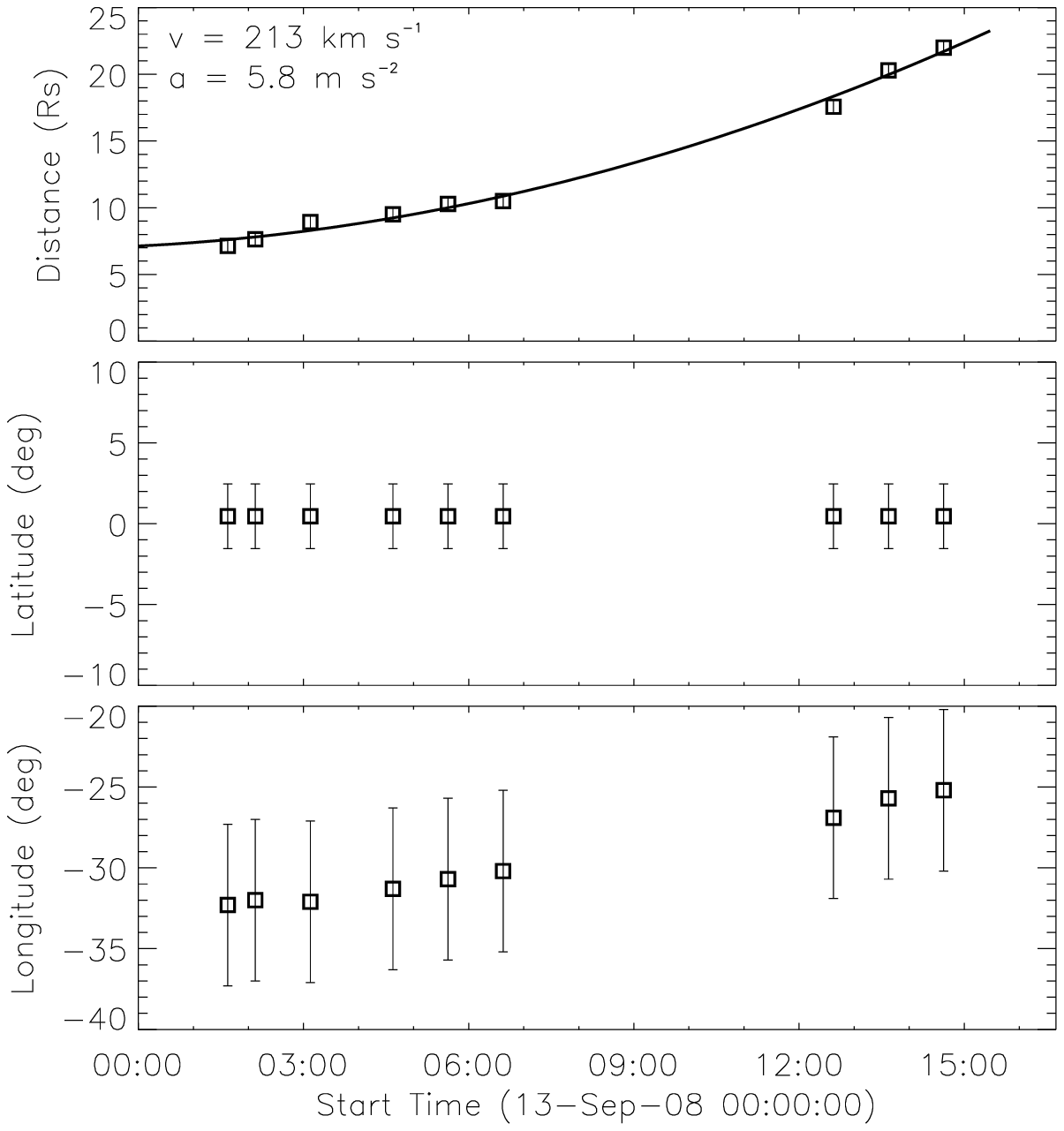}
  \caption{De-projected position of the CME's leading edge derived by GCS model. The errors in distance,
  latitude and longitude are about $0.5R_S$, $2^\circ$ and $5^\circ$, respectively. A systematically westward deflection
is well revealed in the lower panel.}\label{fg_corona}
\end{figure}

Figure~\ref{fg_gcs} shows the CME with modeled flux rope superposed.
In our fitting procedure, the face-on and edge-on widths and
tilt angle are fitted as constants to reduce the degree of freedom.
They are ${78^\circ}_{-18^\circ}^{+34^\circ}$,
${17^\circ}_{-5^\circ}^{+8^\circ}$ and $-11^\circ\pm22^\circ$,
respectively. The errors are estimated following the method by
\citet{Thernisien_etal_2009}, each of which will cause the 10\%
decrease of the best fit. With this configuration, the longitudinal
extent of the CME in the ecliptic plane is estimated to be about
$60^\circ$. The other three parameters are all time- or
distance-dependent as shown in Figure~\ref{fg_corona}. The errors in
the distance, latitude and longitude are about $0.5 R_S$, $2^\circ$
and $5^\circ$, respectively.

Since the GCS model requires that the CME is clearly visible in both
images from STA and STB, the first result is obtained for time at
01:37 UT on September 13, when the CME's leading edge already
reached $7.1R_S$. At that time, the propagation direction of the CME
is about $0.5^\circ$ in latitude and $-32^\circ$ in longitude in the
heliocentric coordinates, i.e., about $2^\circ$ on the west of the
Sun-STB line. The result is in agreement with our previous estimate
of the CME initial propagation direction in Sec.\ref{sec_imaging}.
Furthermore, we have tested the goodness-of-fit by assuming the CME
propagated along the Sun-Earth line, which means that the CME was not 
deflected in interplanetary space. But the fitting result becomes 
much worse.

During the next 13 hours, the CME traveled from $7.1R_S$ to $22R_S$
with an average velocity of about 213 km s$^{-1}$. It experienced an
acceleration process. The acceleration is about 5.8 m s$^{-2}$.
During the period, the latitude of the CME direction did not change,
but the longitude monotonically increased from $-32^\circ$ to
$-25^\circ$. The CME was deflected toward the west by about
$7^\circ$ in the corona. At 15:00 UT when the CME was $22R_S$ away
from the Sun, the CME speed was accelerated to 353 km s$^{-1}$, and
its propagation direction was changed to $9^\circ$ on the west of
the Sun-STB line or $25^\circ$ on the east of the Sun-Earth line.
The CME main body still tends to hit STB rather than the Earth,
which is inconsistent with the in-situ observations presented in the
last section. Thus the CME should be continuously deflected in
interplanetary space.

\subsection{In interplanetary space}\label{sec_ippro}

\begin{figure}[tbh]
  \centering
  \includegraphics[width=\hsize]{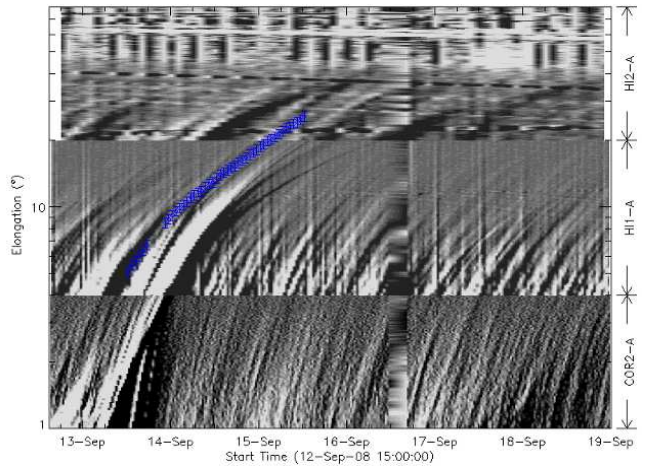}
  \caption{J-map of ecliptic plane generated from COR2, HI1 and HI2 imaging data from STA.
Blue diamonds with error bars indicate the track of the CME's leading edge viewed by STA.
}\label{fg_jmapa}
\end{figure}

\begin{figure}[tbh]
  \centering
  \includegraphics[width=\hsize]{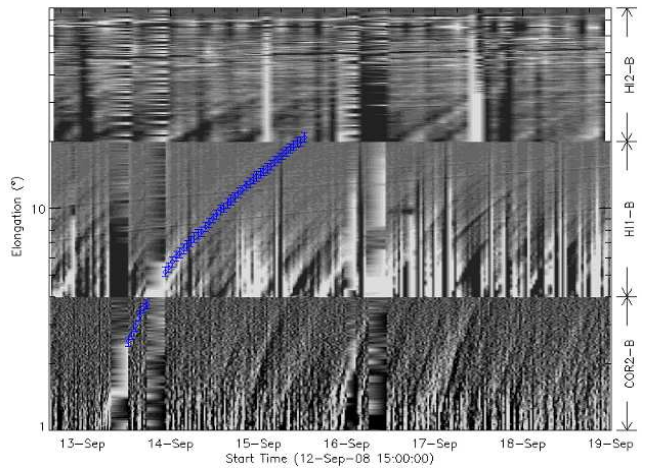}
  \caption{J-map from STB.}\label{fg_jmapb}
\end{figure}

In order to track the CME in interplanetary space, an
elongation-time map, known as J-map \citep{Davies_etal_2009}, is
used. Figure~\ref{fg_jmapa} and \ref{fg_jmapb} show the J-maps
constructed based on the imaging data from COR2, HI1 and HI2 imagers
onboard STA and STB by placing a slice along ecliptic plane. They
provide the information from the corona to 1 AU. Any stripes with a
positive slope in a J-map indicate a featured element moving away
from the Sun. By comparing the stripes in the J-maps with the CME
features in the images from COR2 and HI1, we may locate which one
corresponds to the track of the CME's leading edge in the FOVs of
COR2 and HI1 in both J-maps, and then follow the track into the FOV
of HI2 in the J-maps (as marked by blue diamonds). Since the CME's
leading edge appears weaker and more diffusive with increasing
elongation angle, we set a reasonable error of about $\pm5\%$ of
elongation angle for the measurements. As seen in
Figure~\ref{fg_jmapa} and \ref{fg_jmapb}, the error bars can cover
the width of the diffusing tracks.

Knowing the positions of STA and STB and the elongation angles of
the CME's leading edge measured from the two vantage points, we are
able to derive the heliocentric distance and propagation direction
of the CME with some assumptions. Here two widely-used triangulation
methods are employed. One is the simplest triangulation, in which it
is assumed that the tracks in the two J-maps describe the trajectory
of the same plasma element \citep{Liu_etal_2010a}. The other one is
called harmonic-mean (HM) triangulation, and it assumes that the CME
is a sphere tangent to the solar surface, and the tracks in both
J-maps are on the circular front but are not the same part of the
CME \citep{Lugaz_etal_2009}.

\begin{figure}[tbh]
  \centering
  \includegraphics[width=\hsize]{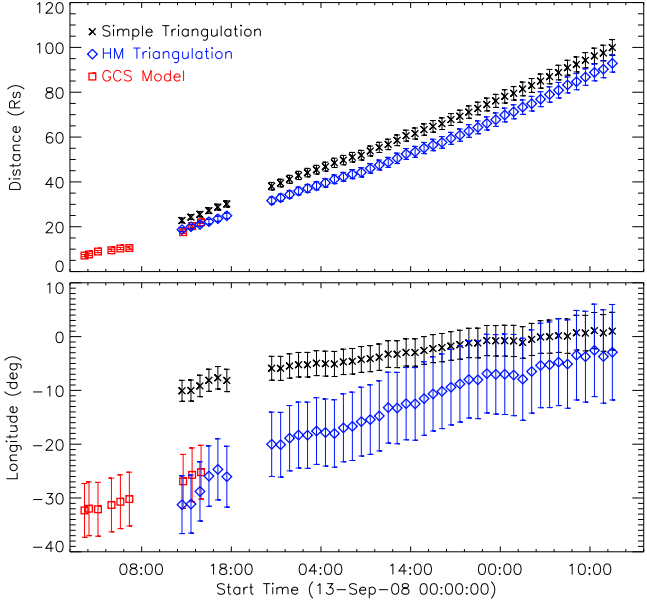}
  \caption{De-projected position of the CME's leading edge derived by triangulation
methods based on the J-maps, showing the entire propagation process of the CME from
the corona to interplanetary space. For completeness, the GCS model results are also
plotted.}\label{fg_tri}
\end{figure}

Figure~\ref{fg_tri} shows the results from the two triangulation
methods, The GCS model results are also plotted for comparison. Both
triangulations show an evident westward deflected propagation of the
CME in interplanetary space even if the uncertainties are taken into
account. The heliocentric distances derived by HM triangulation are
almost the same as those by GCS model for the last three
measurements in COR2 FOV, corresponding to the period September 13
12:00 -- 15:00 UT. The distances derived by the simple triangulation
is systematically larger by $5-10R_S$. The longitude of the
propagation direction derived by HM triangulation is about
$-31^\circ$ at around 12:00 UT, and quickly increased to about
$-25^\circ$ before the CME escaped from the FOV of COR2. The values
are close to those given by GCS model. But the simple triangulation
suggests that the longitude of the CME direction is about
$-10^\circ$ in the FOV of COR2. If this is true, the CME should look
more likely halo in the view from SOHO than in the view from STB,
which is inconsistent with the imaging data presented in
Sec.\ref{sec_imaging}. Thus for this case, HM triangulation gives
more reasonable results. According to the assumptions of the simple
triangulation method, it is expected to be applicable for CMEs with
small extent in longitude. The longitudinal extent of the CME of
interest is about $60^\circ$ (see Sec.\ref{sec_corona}), which is
probably too large to make the recorded tracks in both J-maps being
the same part of the CME.

With the aid of the HM triangulation, it is suggested that the CME
is continuously deflected in interplanetary space. The propagation
longitude changed from $-25^\circ$ at around 15:00 UT to about
$-3^\circ$ at 12:30 UT on September 15 when the CME's leading edge
reached about $93R_S$. In other words, the CME was deflected toward
the west by about $22^\circ$ in $\sim46$ hours or in $71R_S$.
Obviously, the amount of the deflected angle in interplanetary space
is much larger than that in the corona, suggesting that
interplanetary space is a major region where the CME deflection
takes place.

\begin{figure}[tbh]
  \centering
  \includegraphics[width=\hsize]{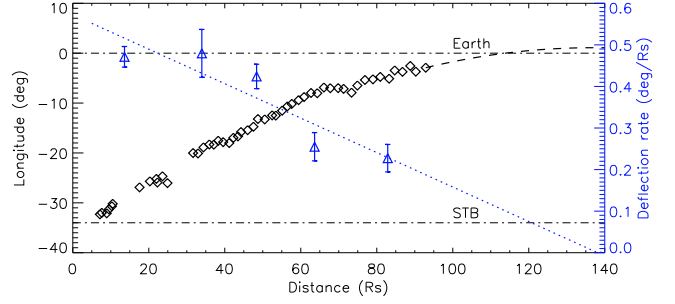}
  \caption{Longitude of the CME's leading edge as a function of heliocentric distance is shown
as diamonds. The deflection rate, i.e., deflection angle per unit distance, which is calculated
on every 10 data points and scaled by the vertical axis on the right, is indicated by blue
triangles. The blue dotted line is the linear fitting to the deflection rate, and dashed line
shows the expected longitudes derived based on the dotted line.}\label{fg_def}
\end{figure}

The deflection rate gradually decreased from the corona to
interplanetary space. Figure~\ref{fg_def} show the longitude of the
propagation direction as a function of the heliocentric distance as
well as the deflection rate. Here, before 15:00 UT on September 13,
we choose the data points from GCS model, and after then, we choose
the data points from HM triangulation. Meanwhile, we divide all the
data points into 5 groups, in each of which there are at least 10
data points, to calculate the deflection rate. The error of the
deflection rate is derived from the linear fitting to the data
points in each group (indicated by the error bars in
Fig.\ref{fg_def}). It is found that the deflection rate is close to
$0.5^\circ/R_S$ before the CME arrived at $40R_S$, and then
gradually dropped below $0.3^\circ/R_S$. Obeying this trend, the
deflection rate will approach zero sooner or later. Simply, we use a
linear fitting to extrapolate the deflection rate and the resultant
propagation longitude, as indicated by the lines in
Figure~\ref{fg_def}. The deflection will probably cease at around
$140R_S$, where the longitude of the CME's propagation direction is
about $1^\circ$. A similar deflection in the interplanetary space
could be found in the paper by \citet{Lugaz_etal_2010}.

These results are highly consistent with the observations. Recall
that the positions of STA and STB in the ecliptic plane are
$+39^\circ$ and $-34^\circ$ away from the Earth, respectively. The
CME initially propagated along the longitude of about $-32^\circ$,
and was finally deflected to about $1^\circ$, which made the CME
being $38^\circ$ away from STA and $35^\circ$ from STB. 
Thus it is possible that STB observed the CME but STA
did not.

\section{Conclusions}\label{sec_con}
In summary, by combining the data from multiple points, i.e., STA,
STB, SOHO and Wind spacecraft, we studied in details the propagation
process of the 2008 September 12 CME from the corona to 1 AU. The
analysis definitely reveals that the CME experienced a westward
deflection throughout the heliosphere. The deflection angle reaches
as large as about $30^\circ$, among which a $20^\circ$-deflection
occurred in interplanetary space. During the period of interest,
there was no other ejection with a similar direction before or after
the CME, suggesting that a CME could be significantly deflected by
solar wind and interplanetary magnetic field.

\begin{figure}[b]
  \centering
  \includegraphics[width=\hsize]{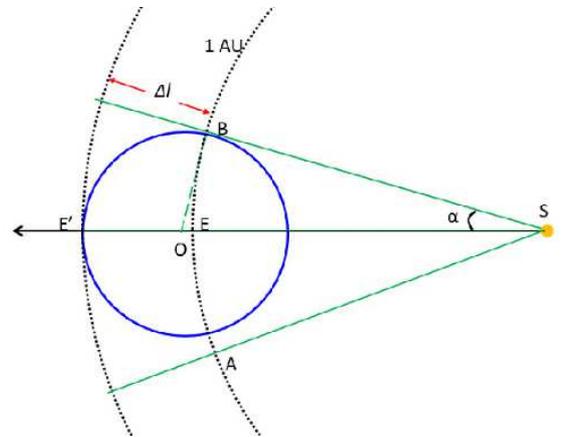}
  \caption{A diagram illustrates the time delay of CME arrival due to its circular-like
front.}\label{fg_delay}
\end{figure}

Deflection not only affects which target will be hit, but also when
a target will be hit. For the CME of interest, we may further
approximate its front in the ecliptic plane to be a circle. As
derived in Sec.\ref{sec_ippro} that the propagation longitude of the
CME at 1 AU almost coincides with the Sun-Earth line, and the CME
flank glanced over STB, we may have the configuration of the CME and
spacecraft as shown in Figure~\ref{fg_delay}. When the CME's flank
arrives at STB, the Earth already dipped in the CME. The time
difference between the arrivals of STB and the Earth is from the
length difference, $\Delta l$, between $\overline{SE'}$ and
$\overline{SB}$. The angle $\alpha$ between the two lines is about
$35^\circ$, and thus $\Delta l$ is about 0.9 AU. Considering the
propagation speed of the CME observed by Wind is about 415 km
s$^{-1}$, the time delay will be as large as 3.7 days. Such delays
were discussed in \citet{Mostl_Davies_2013}. Actually, the CME
cross-section may not be a circle but a ellipse or in a `pancake'
shape \citep[e.g.,][]{Riley_Crooker_2004, Owens_etal_2006}, and the
time delay will be smaller than that derived based on a circle
assumption. Generally, the time delay derived above is consistent
with the in-situ observations, which suggest a 2-day delay.

\begin{figure}[tb]
  \centering
  \includegraphics[width=\hsize]{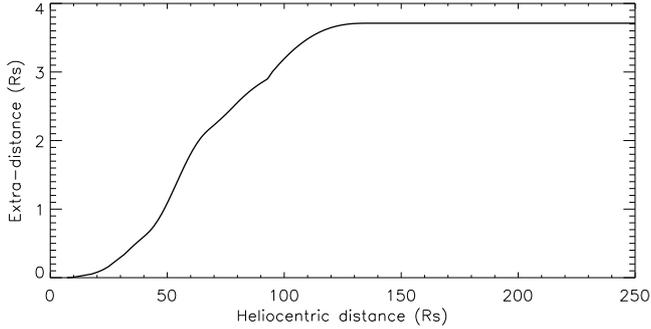}
  \caption{Extra-distance the CME propagating through due to its curved
trajectory. It is derived based on the deflection angle shown in
Fig.\ref{fg_def}.}\label{fg_dis}
\end{figure}

Thus, in the prediction of CME arrival time, the deflection combined
with the CME geometry are important factors that need to be taken
into account. Besides, as a consequence, the curved trajectory,
along which a deflected CME actually propagates, is a minor factor
influencing the accuracy of a prediction. Obviously, the length of a
curved trajectory must be longer than a straight trajectory.
Figure~\ref{fg_dis} shows the length difference between the curved
trajectory and the heliocentric distance of this event. The length
difference, what we call extra-distance here, is about $4R_S$.
The average transit time of the CME is about 100
hours, corresponding to an average transit speed of about 400 km
s$^{-1}$. If the $4R_S$ extra-distance was not taken into account,
the prediction of the CME arrival time based on some simple
empirical model will have a two-hour error, which is less than but
on the same order of the typical error for CME arrival time
prediction.

From the above analysis, some outstanding questions about space
weather forecasting emerge. What is the cause of the CME deflection
and how to precisely predict the trajectory of a CME? In the
following section, we will discuss the mechanism of CME deflection
in interplanetary space.

\section{Discussion on the mechanism of the CME deflection}\label{sec_cause}

It is well accepted that the deflection of a CME in the corona,
usually within $5-10R_S$, is controlled by gradient of magnetic
energy density \citep{Shen_etal_2011, Gui_etal_2011,
Kahler_etal_2012, Zuccarello_etal_2012, Kay_etal_2013}. Generally,
the magnetic energy density reaches the minimum at heliospheric
current sheet, which locates near ecliptic plane during solar
minima. That is why CMEs in solar minima tend to propagate toward
the ecliptic plane \citep{Cremades_Bothmer_2004, Wang_etal_2011}. Is
it also the cause of the CME deflection in interplanetary space? To
answer the question, we investigate the magnetic field and current sheet at
1 AU.

The magnetic field in interplanetary space is obtained by utilizing
a 3-dimensional MHD numerical method, in which a
corona-interplanetary total variation diminishing (COIN-TVD) scheme
is adopted \citep{Feng_etal_2003, Feng_etal_2005}. Starting from
Parker's solar wind solution and the potential magnetic field
extrapolated from magnetic field distribution at photosphere
observed by WSO during the Carrington rotation 2074 covering the
period of the CME, we use this scheme to establish a steady state of
background solar wind and interplanetary magnetic field. A detailed
description of the scheme and its application can be found in our
previous work \citep{ShenF_etal_2007, ShenF_etal_2009,
ShenF_etal_2011a, ShenF_etal_2013}.

\begin{figure}[tb]
  \centering
  \includegraphics[width=\hsize]{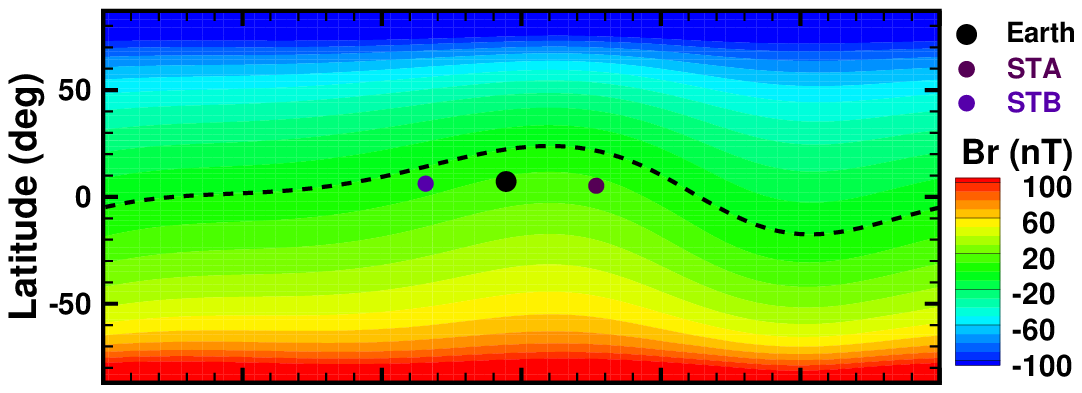}\\[8pt]
  \includegraphics[width=\hsize]{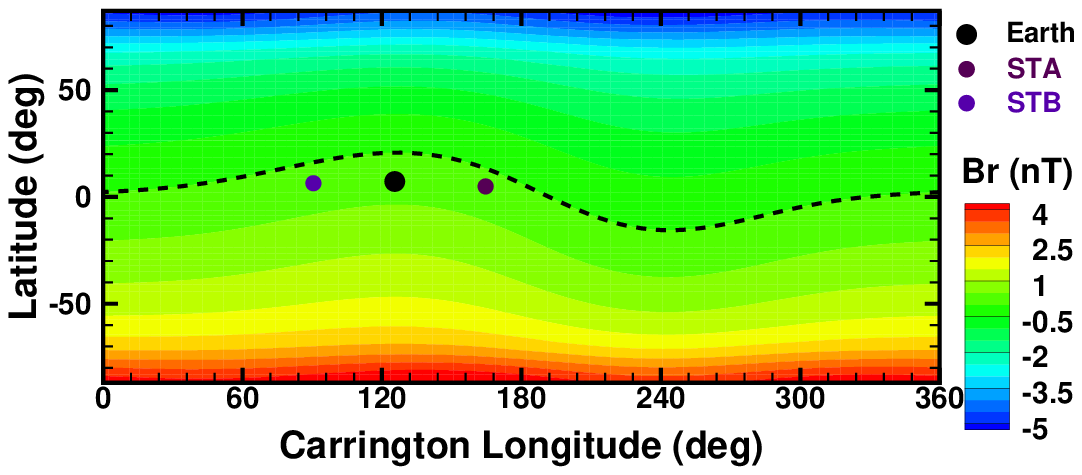}
  \caption{Carrington maps of radial component of magnetic field at $20R_S$ (upper panel)
and 1 AU (lower panel), respectively, which are
constructed by a 3-dimensional MHD simulation. Dashed lines indicate the
location of heliocentric current sheet. The dots on the two maps show the positions
of STA, STB and the Earth at 13:30 UT on September 13 and 04:40 UT on September 17,
when the CME arrived at $20R_S$ and $1 AU$, respectively.}\label{fg_hcs}
\end{figure}

Figure~\ref{fg_hcs} shows the distribution of radial-component of
magnetic field at $20R_S$ and 1 AU. The location of current sheet is
indicated by the dashed lines. The positions of STA, STB and the
Earth projected on the maps are marked as dots. Note that the CME
arrived at $20R_S$ and 1 AU around 13:00 UT on September 13 and
04:30 UT on September 17, respectively. Thus the Carrington
longitudes of these positions are different between the two maps. It
is clear that the CME should deflect toward the north rather than
the west if it still obeyed the deflection law in the corona. This
opposite result suggests that there should be other causes of the
deflection in interplanetary space.

\begin{figure*}[!tbh]
  \centering
  \includegraphics[width=0.510\hsize]{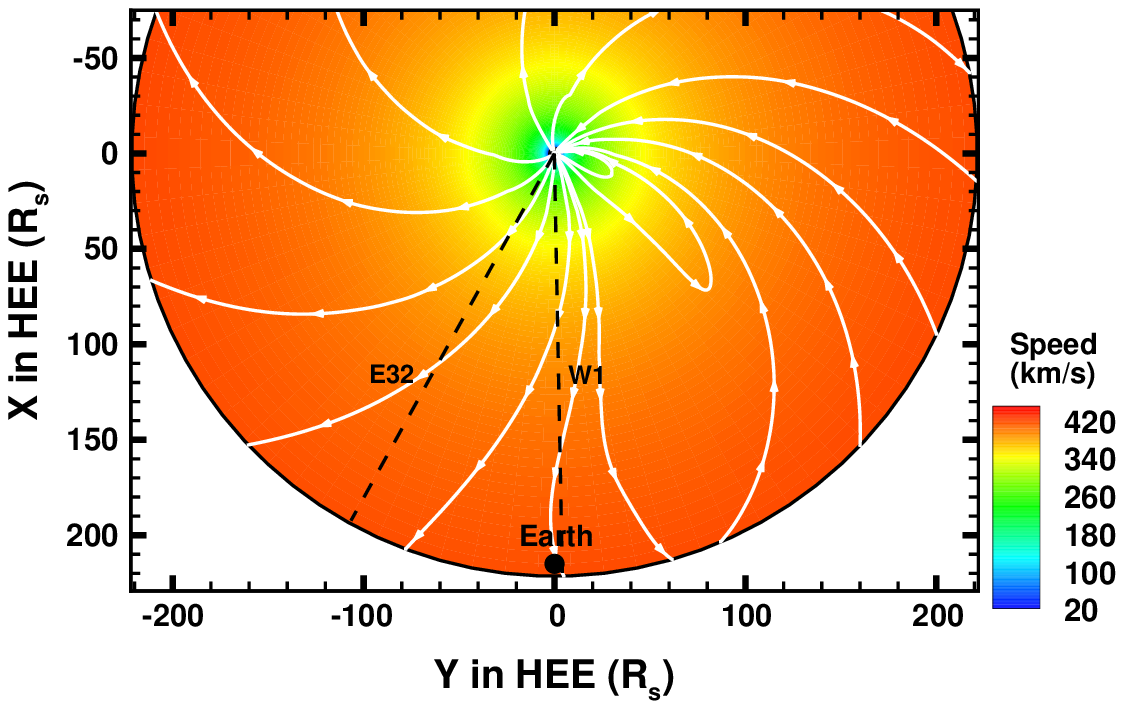}
  \includegraphics[width=0.480\hsize]{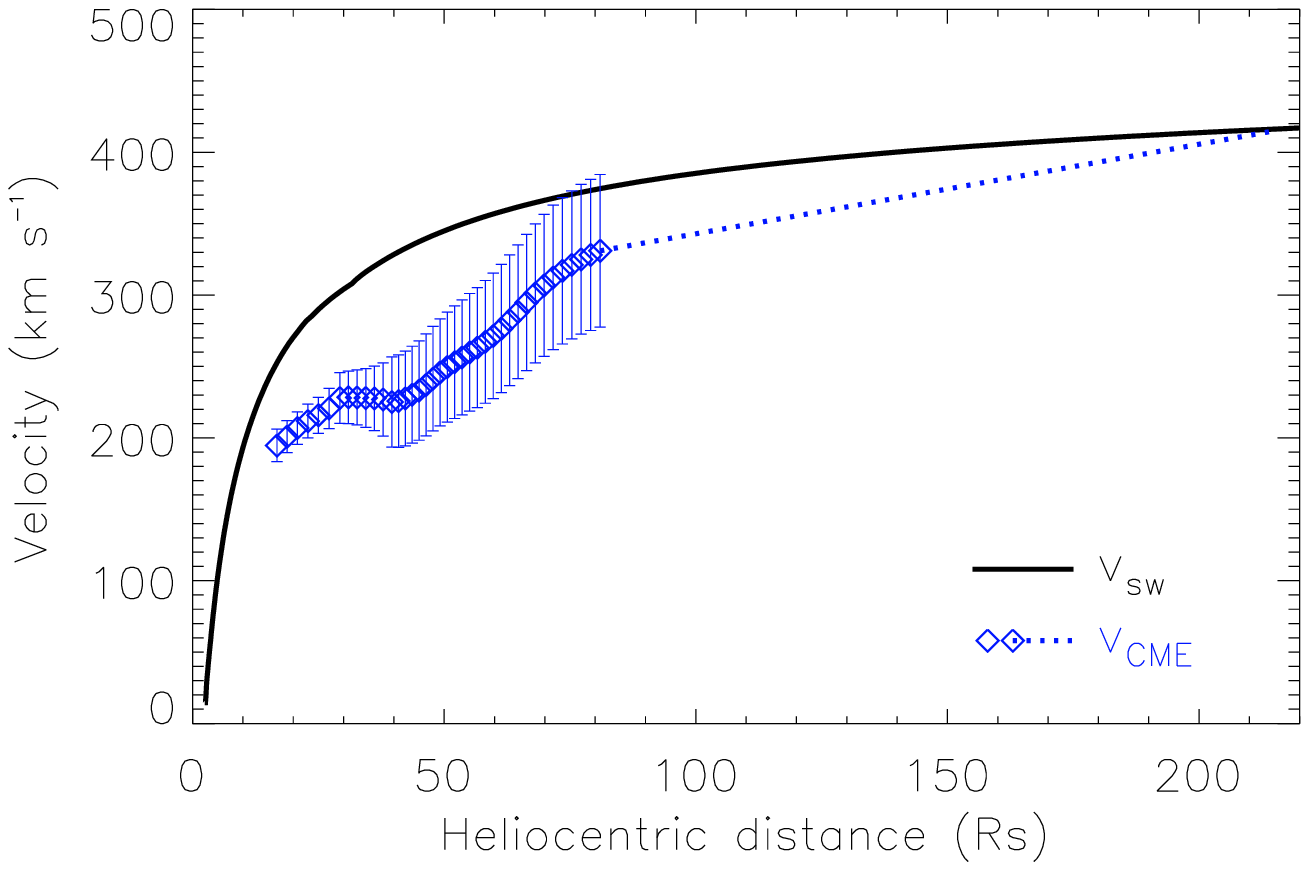}
  \caption{Left panel shows the solar wind velocity distribution in the ecliptic plane derived
by 3-dimensional MHD simulation, and the right panel gives the
average velocity of the solar wind within longitude of $-32^\circ$
and $1^\circ$ versus the heliocentric distance. As comparison, the
CME speed is plotted as the blue diamonds and dashed line (see main
text for details). The white lines in the left panel are the
magnetic field lines.}\label{fg_sw}
\end{figure*}

In the corona, magnetic field is dominant as solar wind has not
fully developed. Thus the magnetic energy density gradient guides
the trajectory of a CME in the corona. However, in interplanetary
space, magnetic field drops quickly and solar wind becomes dominant.
Thus interaction between the CME and background solar wind should be
the cause of the CME deflection in interplanetary space. A decade
ago, \citet{Wang_etal_2004b} proposed a kinematic
model to describe the CME's trajectory modulated by the interplanetary
magnetic field carried by solar wind. In that model, the
magnetic field is simply supposed to be strong enough to ensure that a 
CME follows the Parker spiral. It predicts that a slow CME will be
deflected toward the west due to faster solar wind accumulating
behind the CME and overtaking it from the east, while a fast CME will 
be deflected toward the east due to slower solar wind being piled up ahead of 
the CME and overtaken from the west \citep[ref. to Fig.4 in][]{Wang_etal_2004b}. 
The model is so far the only theoretical
model (except those MHD numerical simulation models) to describe the
deflection of CMEs in the ecliptic plane. It is interesting to see
how well the model could reproduce the trajectory of the isolated
CME launched on 2008 September 12.

The physical picture of the model is similar to the solar wind deflection 
within a CIR, that the preceding slow solar wind has a deflection toward the 
west and the following fast solar wind toward the east \citep[e.g.,][]{Siscoe_etal_1969, Gosling_Pizzo_1999, 
Broiles_etal_2012}. 
An apparent difference between them is that the deflection 
speed of the solar wind within CIRs is still significant at 1 AU 
\citep{Broiles_etal_2012}, but that of the CME is not (Fig.\ref{fg_def}).
This is because the CME speed approaches the solar wind speed at 1 AU.
Thus we think that the velocity difference between two interacting system
is the essential cause of the two phenomena.

The basic formula to describe the deflection of a CME
is the second equation in Eq.5 of \citet{Wang_etal_2004b}. Only the
radial components of the CME velocity and background solar wind
velocity, namely $v_r$ and $v_{sw}$, are required. It should be
noted that the equation was developed for constant CME speed and
constant solar wind speed. To accept varied speeds, we just need to
slightly change the equation to the following one
\begin{eqnarray}
d\phi=\Omega\left(\frac{1}{v_r}-\frac{1}{v_{sw}}\right)dr\label{eq_1}
\end{eqnarray}
in which $\phi$ is the longitude, $r$ the heliocentric distance and
$\Omega$ the angular speed of the Sun's rotation. For
completeness, a derivation of the above equation, different but much
briefer than that in \citet{Wang_etal_2004b}, is given in the
Appendix. The total change of longitude of a CME is the integral of
Eq.\ref{eq_1}, which could be numerically calculated once we know
$v_{sw}$ and $v_r$, which are both functions of heliocentric
distance $r$.

To obtain the solar wind speed, $v_{sw}$, we utilize the MHD
numerical method again. The left panel of Figure~\ref{fg_sw} shows
the radial component of the solar wind velocity in ecliptic plane
with interplanetary magnetic field lines superimposed. Since the
previous analysis have suggested that the CME was deflected from
$-32^\circ$ to $1^\circ$, we use the averaged solar wind speed
between the two longitudes. The profile of the speed is shown in the
right panel of Figure~\ref{fg_sw}. Within first $20R_S$ the
background solar wind quickly accelerates to 300 km s$^{-1}$, and
then gradually accelerates to more than 410 km s$^{-1}$ at 1 AU.

For comparison, the CME bulk propagation speed, $v_r$, is plotted as
blue diamonds and a dashed line in the right panel of
Figure~\ref{fg_sw}. The speed below $90R_S$ is derived from the
height time plot of the CME leading edge in Figure~\ref{fg_tri} with
expansion speed deducted. Here we assume that the CME expanded with
a constant angular width. At 1 AU, Wind data suggests that the CME
expansion speed is about 45 km s$^{-1}$, the bulk propagation speed
is about 415 km s$^{-1}$ and the radius is about 0.14 AU. It means
that the bulk propagation speed of the CME is 11\% smaller than the
speed of the CME leading edge, and the heliocentric distance is 14\%
shorter than that of the leading edge. The speed beyond $90R_S$ is
simply obtained by a linear extrapolation.

\begin{figure}[tbh]
  \centering
  \includegraphics[width=\hsize]{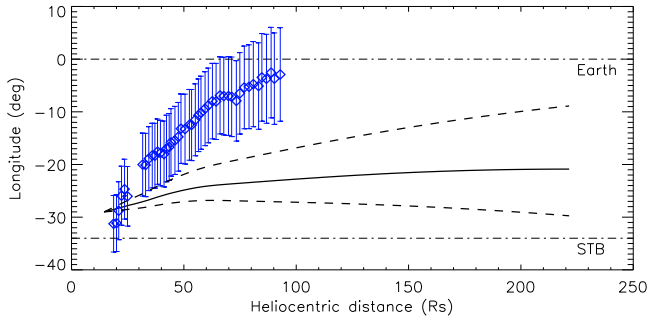}
  \caption{Observed (blue diamonds with error bars) and model
predicted (black lines) longitudes of the CME. The two dashed lines
indicate the predicted longitudes by considering $\pm15\%$ error in
the CME velocity. Refer to Sec.\ref{sec_cause} for
details.}\label{fg_traj}
\end{figure}

By inputting $v_{sw}$ and $v_r$ into our model, we find that the
propagation longitude of the CME changes about $8^\circ$ from the
initial value of $-29^\circ$ to $-21^\circ$, as shown by the solid
line in Figure~\ref{fg_traj}. The amount of the deflection predicted
by the model is much smaller than that derived from
observations. Considering that deflection in our model is
substantially due to the difference between $v_r$ and $v_{sw}$, we
could expect that the error in either velocity will cause the error
in deflection angle. The two dashed lines show the different CME
trajectory if the CME velocity was 15\% higher (the lower line,
suggesting a smaller deflection) or lower (the upper line,
suggesting a larger deflection). The latter suggests a
$20^\circ$ deflection, closer to but still smaller than the
observations.

In summary, the deflection of a CME in interplanetary space
has a different cause of that in the corona. Although our
kinematic model predicts a westward deflection of the CME
originating on September 12, the modeled trajectory is not good
enough. The deviation between the model predicted and the observed
trajectory could be from the highly-ideal assumptions used in our
model and/or other unknown factors/processes that take place during
the solar wind-CME interaction. For example, we only consider the Parker spiral 
magnetic field lines shaped by solar wind but do not fully take the kinetic 
energy carried by the solar wind into account. But in interplanetary
space, the kinetic energy should be stronger than magnetic energy.

\begin{acknowledgments}
The data for this work are available at the official websites of
STEREO, SOHO and Wind spacecraft. We acknowledge the use of them.
STEREO is the third mission in NASA's Solar Terrestrial Probes
programme, and SOHO is a mission of international cooperation
between ESA and NASA. We thank anonymous referees for valuable
comments and suggestions. This work is supported by grants from MOST
973 key project (2011CB811403 and 2012CB825600), CAS (Key Research
Program KZZD-EW-01 and 100-Talent Program), NSFC (41131065,
41121003, 41274173, 41074121 and 41174150), MOEC (20113402110001)
and the fundamental research funds for the central universities.
N.~L. was partially supported by NSF grant AGS-1239704.
\end{acknowledgments}

\appendix
\section{Derivation of kinematic model of the CME deflection}

\begin{figure}[tbh]
  \centering
  \includegraphics[width=\hsize]{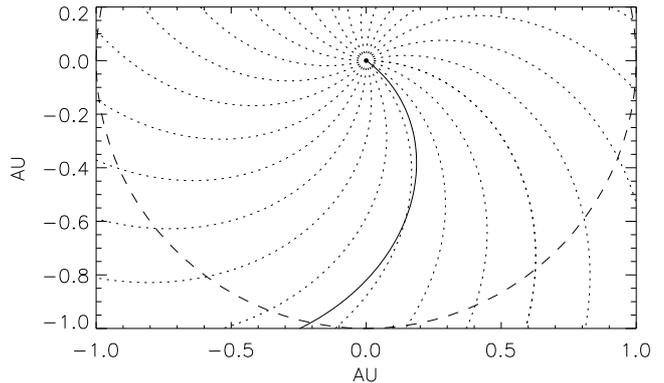}
  \caption{Illustration of the model.}\label{fg_imf}
\end{figure}

The model assumes that the background solar wind and interplanetary
magnetic field (IMF) is dominant and the CME is a fluid parcel, so
that the CME in the ecliptic plane tends to move following IMF
lines. Figure~\ref{fg_imf} illustrates the model. Dotted lines are
Parker spiral magnetic field lines, and the solid line is the
unaffected field line connecting to the CME which moves radially
with a speed slower than the background solar wind. Since magnetic
field lines cannot cross over each other, the slow CME should be
deflected toward the west to make the solid line coinciding with the
spiral magnetic field line starting from the same place on the Sun
(or the IMF lines will be deformed if CME kinetic energy was
dominant). Thus the model is more suitable for slow CMEs.

The detailed derivation of the model was given in
\citep{Wang_etal_2004b}. Here, we present another approach to derive
the model. It is a 2-D model in the ecliptic plane. Let $v_{sw}$ be the 
solar wind speed, $\Omega$ the solar rotation, $\varphi_i$ the initial longitude
and $t$ the time since the plasma element left the Sun, then
the Parker spiral IMF shown in Figure~\ref{fg_imf} is given by
\begin{eqnarray}
r_{0}&=&v_{sw}t \label{eq_02}\\
\varphi_{0}&=&\varphi_i-\Omega t
\end{eqnarray}
Assuming that the CME is a plasma parcel with a radial speed of $v_r$,
the magnetic field line drawn by the CME
is given by
\begin{eqnarray}
r&=&v_rt\\
\varphi&=&\varphi_i-\Omega t+\Delta\phi(t)\label{eq_03}
\end{eqnarray}
which should satisfy the following condition because the CME is assumed to follow the Parker spiral of the IMF,
\begin{eqnarray}
&&\frac{r_0}{\varphi_0-\varphi_i}=\frac{r}{\varphi-\varphi_i}
\end{eqnarray}
Here $\Delta\phi$ is the time-dependent or distance-dependent
deflection angle of the CME. It is easy to derive that
\begin{eqnarray}
\Delta\phi(t)=\frac{v_{sw}-v_r}{v_{sw}}\Omega t
\end{eqnarray}
or
\begin{eqnarray}
\Delta\phi(r)&=&\frac{v_{sw}-v_r}{v_{sw}v_r}\Omega r \nonumber\\
&=&\left(\frac{1}{a}-\frac{1}{a_0}\right) r \label{eq_01}
\end{eqnarray}
in which $a=v_r/\Omega$ and $a_0=v_{sw}/\Omega$. Eq.\ref{eq_01} is
exactly the same as Eq.5 in \citet{Wang_etal_2004b}.

The above derivation uses the constant velocity for both solar wind
and the CME. To accept varied velocity, we just convert
Eq.\ref{eq_02}--\ref{eq_03} to the differential form
\begin{eqnarray}
&&\left\{\begin{array}{l}
dr_0=v_{sw}dt\\
d\varphi_0=-\Omega dt
\end{array}\right.\\
&&\left\{\begin{array}{l}
dr=v_{r}dt\\
d\varphi=-\Omega dt+d\phi(t)
\end{array}\right.
\end{eqnarray}
Then we can get the deflection angle
\begin{eqnarray}
d\phi&=&\frac{v_{sw}-v_r}{v_{sw}}\Omega dt \nonumber\\
&=&\left(\frac{1}{v_r}-\frac{1}{v_{sw}}\right)\Omega dr
\end{eqnarray}
as well as the angular velocity of the CME
\begin{eqnarray}
\omega=\frac{d\phi}{dt}=\frac{v_{sw}-v_r}{v_{sw}}\Omega
\end{eqnarray}

The interaction between solar wind and CMEs will not only affect the
angular motion but also the radial motion, i.e.,
acceleration/deceleration, of the CME. The model can only predict
the change of angular motion. The change of the radial motion caused
by the solar wind interaction has been taken into account by
adopting changing $v_r$ of the CME as derived from the heliospheric observations.

\bibliographystyle{agufull}
\bibliography{../../ahareference}

\end{document}